# Nitrogen-Doped Graphene Sheets Grown by Chemical Vapor Deposition: Synthesis and Influence of Nitrogen Impurities on Carrier Transport

Yu-Fen Lu[a,b,#], Shun-Tsung Lo[c,#], Jheng-Cyuan Lin[b], Wenjing Zhang[a], Jing-Yu Lu[b], Fan-Hung Liu[c], Chuan-Ming Tseng[d], Yi-Hsien Lee[e], Chi-Te Liang[b,c], and Lain-Jong Li[a]

[a]Institute of Atomic and Molecular Sciences, Academia Sinica, Taipei, 106, Taiwan
[b]Department of Physics, National Taiwan University, Taipei 106, Taiwan
[c]Graduate Institute of Applied Physics, National Taiwan University, Taipei 106, Taiwan
[d]Institute of Physics, Academia Sinica, Taipei 115, Taiwan
[e]Department of Materials Science and Engineering, National Tsing Hua University, Hsinchu 300, Taiwan
# These authors contributed equally.
E-mail: ctliang@phys.ntu.edu.tw and lanceli@gate.sinica.edu.tw

**Abstract**

A significant advance toward achieving practical applications of graphene as a two-dimensional material in nanoelectronics would be provided by successful synthesis of both n-type and p-type doped graphene. However reliable doping and a thorough understanding of carrier transport in the presence of charged impurities governed by ionized donors or acceptors in the graphene lattice are still lacking. Here we report experimental realization of few-layer nitrogen-doped (N-doped) graphene sheets by chemical vapor deposition of organic molecule 1, 3, 5-triazine on Cu metal catalyst. By reducing the growth temperature, the atomic percentage of nitrogen doping is raised from 2.1 % to 5.6 %. With increasing doping concentration, N-doped graphene sheet exhibits a crossover from p-type to n-type behavior accompanied by a strong enhancement of electron-hole transport asymmetry, manifesting the influence of incorporated nitrogen impurities. In addition, by analyzing the data of X-ray photoelectron spectroscopy, Raman spectroscopy, and electrical measurements, we show that pyridinic and pyrrolic N impurities play an important role in determining the transport behavior of carriers in N-doped graphene sheets.

Graphene, a monolayer of sp$^2$-bonded carbon atoms, has been attracting much interest due to its unique band structure.[1-4] Conically shaped conduction and valence bands (or valleys), touching at two nonequivalent corners of the hexagonal Brillouin zone called Dirac points, make charge transport in graphene remarkably different from that of conventional two-dimensional (2D) systems.[5, 6] With the presence of scattering mechanism that couples such two independent valleys (intervalley scattering), localization effects can occur.[7] Without it or in the presence of only intravalley scattering, the transport carriers would not be impeded by disorder potential (the well celebrated Klein tunneling effect),[8] leading to the absence of backscattering under which the graphene mobility can be extremely high. Intervalley and intravalley scattering are induced by short-range scatterers (*e.g.*, from lattice defects) and long-range scatterers (*e.g.*, from ionized impurities or ripples), respectively.[5, 6] Therefore transport behavior in graphene is strongly relevant to the nature of disorder. A wide variety of exciting physical phenomena including ambipolar electric field effect and absence of backscattering render graphene a promising material for use in nanoelectronics. For pristine graphene without doping, the Fermi level is precisely at the Dirac point. To date, considerable efforts have been focused on preparing either n-type or p-type doped graphene so as to fully realize graphene-based electronics. The n-type doping shifts the Fermi level above the Dirac point whereas the p-type doping shifts the Fermi level below it. Therefore, it is highly desirable to find a controllable way to tailor the electrical properties of graphene.

Molecular charge-transfer doping with different kinds of absorbents like NH$_3$ as electron donor[1, 9] and NO$_2$ as electron acceptor,[10] for instance, is an effective approach to achieve doping in graphene, which is usually applied to decorate graphene surface without deteriorating carrier mobility significantly.[10, 11] In addition, doped graphene can also be synthesized by chemical methods[12-19] through which extrinsic atoms are incorporated into the graphene lattice and thereby the resulting doped graphene would be more stable compared with that from molecular doping. Most studies have focused on nitrogen-doped (N-doped) graphene due to its interesting prospects for various applications in biosensors,[17] field-effect transistors,[20] lithium batteries,[21] supercapacitors,[22] *etc*. A number of approaches have been proposed to synthesize N-doped graphene, such as chemical vapor deposition (CVD) of methane in the presence of ammonia,[12] thermal annealing of graphene oxide in ammonia,[16] and nitrogen plasma treatment of graphene.[17] Both electrical and optical properties of N-doped graphene are demonstrated to be strongly relevant to the bonding configuration. There are predominantly three different bonding characters for the embedded N atoms: graphitic (substitutional), pyridinic, or pyrrolic N. In the graphitic configuration, a fraction of the excess valence electrons from N atoms after forming bonds with three sp$^2$ hybridized carbon neighbors is delocalized and leads to n-type doping of graphene.[23, 24] Instead, according to the calculations based on density functional theory (DFT),[13, 24-26] when either pyridinic or pyrrolic N which refers to sp$^2$ hybridized N bonded with two sp$^2$ hybridized carbon neighbors with adjacent carbon vacancies is dominant, p-type behavior should occur. On the other hand, in chemically derived graphene, increasing doping concentration is ubiquitously accompanied by an increase in charged impurities (from ionized donors or acceptors), which offers an excellent platform for systematically studying the influence of Coulomb scattering from ionized impurities on graphene carrier transport.

In graphene subject to charged-impurity scattering and screening effect,[8] several interesting transport features such as the widely observed linear dependence of conductivity on the applied gate voltage[27, 28] and the mobility asymmetry between electrons and holes are expected to occur.[28] Massless Dirac fermions in graphene are theoretically shown to be scattered more strongly when they are attracted to a charged impurity than when they are repelled from it and thereby electron and hole transport features would be asymmetric with respect to the Dirac point.[29] Such anomalous features have already been investigated in potassium doped graphene using molecular doping method.[30] In addition, they further demonstrate that carrier-density inhomogeneity induced by charged impurities would affect the position of conductivity minimum. However, at the present stage, detailed transport study of synthetically controlled N-doped graphene is still lacking. Here, we report preparation and electrical measurements of CVD-derived few-layer nitrogen-doped graphene sheets with three different doping concentrations using 1, 3, 5-triazine (abbreviated as triazine below) molecules as the sole source of both carbon and nitrogen and Cu foil as a catalyst. With increasing doping concentration *via* decreasing the growth temperature, a crossover from p-type to n-type behavior occurs correspondingly with a strong enhancement of electron-hole transport asymmetry, which indicates that charged-impurity scattering plays a dominant role in the resulting electrical properties of N-doped graphene sheets. Furthermore, we provide experimental evidence revealing the importance of incorporated pyridinic and pyrrolic N dopants. Our experimental findings are crucial for better understanding of intriguing 2D physics in graphene as well as for the development of graphene-based nanoscale devices.

**Results and discussion**

Figure 1a illustrates the conversion of organic molecule triazine into N-doped graphene *via* CVD method by the aid of Cu catalyst, during which the nitrogen atoms are involved with the recombination of triazine molecules and thereby doping of graphene can occur. As demonstrated in the schematic structure of N-doped graphene, doping of graphitic N atoms does not destroy graphene crystal structure significantly[22] preserving the high-quality properties of graphene, whereas the incorporation of pyridinic and pyrrolic N are always accompanied by the appearance of structural defects including bonding disorder and vacancies (short range scatterers) which increases the amount of disorder in the system. In addition, for all three bonding conditions, charged-impurity scattering occurs in the presence of ionized nitrogen impurities (long range scatterers), complicating the physics of 2D transport in graphene.[31-35]

The AFM images and the corresponding cross-section profiles shown in Figs. 2a-2c indicate that all the samples with different growth temperatures consists of few-layer graphite sheets. Figures 3a-3c show the scanning electron microscope (SEM) images of one of the electrical contacts as well as the adjacent graphene sheets grown at 900 ºC, 800 ºC, and 700 ºC, respectively. Such results indicate fairly uniform graphene sheets in our case. Figures 4a-4c show the tunneling electron microscope (TEM) images as well as selective area electron diffraction (SAED) patterns of our N-doped graphene sheets grown at 900 ºC, 800 ºC, and 700 ºC, respectively. The SAED yields ring-shaped pattern consisting of many diffraction spots. This may indicate the otherwise crsytalline graphene sheets became partially misoriented in the N-doped graphene sheets due to structure distortion casued by intercalation of nitrogen atoms into its graphitic planes.[14] This may also be due to stacking of graphene[36] when transferring substrate-free graphene onto TEM grids for measurements.

Figures 5a and 5b show the XPS spectra around C 1s and N 1s peak, respectively, for three different growth temperatures of 900 ºC, 800 ºC, and 700 ºC, confirming the doping of graphene. The C 1s peak can be decomposed into at least two components for all the samples. The strongest peak at ~285 eV corresponds to $sp^2$-hybridized carbon atoms in graphene whereas the side peak at ~288 eV is assigned to $sp^3$ C atoms bonded with N [12]. Moreover, three characteristic peaks centered at 398 eV, 400 eV, and 401 eV, corresponding to pyridinic, pyrrolic, and graphitic N, appear in the N 1s spectrum.[12, 20] It should be mentioned that wide range of binding energies are found in the literature. For example, Wei et al. report a difference of 3.5 eV between graphitic and pyridinic N [12] while a difference of 2.1 eV is found in the seminal work of Jin et al.[15] In our study, it is ~3 eV. The atomic percentage of N (N/C) in our N-doped graphene sheet grown at 900º C, 800 ºC and 700 ºC, estimated by XPS, are about 2.1 %, 4.4 %, and 5.6 %, respectively, suggesting that the doping concentration is raised by reducing the growth temperature. Such results are tabulated in Table 1. Moreover, as can also be found in Table 1, the composition of bonding structure is varied when the growth temperature is different. The atomic percentages of pyridinic and pyrrolic N obtained by multiplying the total N content (N/C) by the corresponding composition ratio increase monotonically with lowering the growth temperature. For all the N-doped graphene sheets, the incorporated nitrogen atoms are mainly in the form of pyridinic and pyrrolic N.

Raman spectroscopy as a powerful tool for identifying carbon materials was used to probe the doping effects in our experiments.[37-39] As shown in Fig. 6a, the main features in the Raman spectra of N-doped graphene grown at 900 ºC are the G band and D band, which lie at around 1592 cm$^{-1}$ and 1374 cm$^{-1}$, respectively. The G band corresponds to optical $E_{2g}$ phonons at the Brillouin zone center. The D band is attributed to the breathing mode of sp$^2$-rings and requires a defect for its activation *via* an intervalley double-resonance (DR) Raman process. Another characteristic feature in graphene, the 2D band, appears at round 2710 cm$^{-1}$. This band is due to the same intervalley DR process but no defects are required for its activation, in contrast to that resulting in the D band. Figures 6a-6c compare the Raman spectra of N-doped graphene sheets grown at different temperatures. The zoom-in view around the 2D band is shown in the insets. For all the samples, high intensity of D band is apparently seen, indicative of the presence of significant defects. In addition, as presented in Fig. 6c for the sample prepared at 700 ºC, we find that the G band is asymmetric, a signature of *D'* band. Defects that cause intravalley DR process are responsible for this. The combination mode D+D′ around 2927 cm$^{-1}$ can be found in all the samples. Here the intensity, position, and width of each characteristic Raman band are quantified by using Lorentzian fits for clarity. (see Supporting Information for detailed fitting results). The defect-related D and D+D′ bands become sharper while the 2D band becomes broader as the growth temperature is reduced, suggesting that N-doped graphene sheets are much more disordered at a reduced growth temperature. With lowering the growth temperature, the intensity ratio of D and G band ($I_D/I_G$), a measure of the amount of defects leading to intervalley scattering,[27, 28] increases. Correspondingly, the 2D to G intensity ratio ($I_{2D}/I_G$) decreases. These ratios are listed in Table. 2. However, it is difficult to use the $I_{2D}/I_G$ ratio in evaluating the amount of disorder and the doping level in N-doped graphene sheets, since both 2D-and G-band features are highly relevant to layer numbers, strains, doping, defects, *etc*.[40] As a consequence, cooperating with XPS data, it is known that increasing pyridinic and pyrrolic N doping with lowering the temperature of CVD growth induces a large amount of disorder mostly due to bonding disorder and vacancies as depicted in Fig. 1.

Pyridinic and pyrrolic nitrogen atoms are different in view of their bonding configurations with carbon atoms. As far as the transport properties are concerned, they may play a similar role since introduction of either pyridinic or pyrrolic nitrogen atoms would induce a large amount of structural defects. However a thorough understanding of the doping effect caused by pyridinic and pyrrolic N impurities are still lacking, which will be the main subject in the following transport studies.

Figures 7a-7c show the conductance as a function of voltage $V_g$ applied on the p-type Si substrate which acts as a bottom gate $G(V_g)$ at three corresponding growth temperatures. These measurements were performed at room temperature. Characteristic ambipolar electric field effect of graphene can be clearly seen. At high $V_g$ away from $V_{g,min}$ where the conductance minimum $G_{min}$ occurs, the conductance is found to depend linearly on $V_g$ in both electron- and hole-transport regimes, implying the significance of charged-impurity scattering in N-doped graphene sheets.[28-30] Therefore the conductance $G$ can be described by

$$G(V_g) = \begin{cases} \dfrac{W}{L} c_g \mu_e (V_g - V_{g,crossing}) + G_{crossing} & V_g > V_{g,crossing} \\ -\dfrac{W}{L} c_g \mu_h (V_g - V_{g,crossing}) + G_{crossing} & V_g < V_{g,crossing} \end{cases} \quad (1)$$

with the channel width $W$ (~ 400 μm) and length $L$ (~ 200 μm), the gate capacitance per unit area $c_g$ (~1.15 10$^{-8}$ F/cm$^2$ for the 300 nm thick SiO$_2$), electron mobility $\mu_e$, hole mobility $\mu_h$, and the residual conductance $G_{crossing}$ occurring at $V_g = V_{g,crossing}$. The crossing of the linear fits denoted by the red lines shown in Figs. 7a-7c defines $G_{crossing}$ and $V_{g,crossing}$. In addition, from such linear fits, carrier mobility can be estimated. At high $V_g$, deviation from linear dependence occurs, which is attributed to the presence of short-range scatterers such as structural defects[28, 33] and some kinds of incorporated chemical dopants.[24, 29] On the other hand, the Dirac point position usually assigned to $V_{g,min} \sim V_{g,crossing}$ moves from positive to negative value (from 2.1 V to -1.1 V), revealing a transition from p-type to n-type behavior, when the growth temperature is reduced from 900 ºC to 800 ºC. However, at the growth temperature of 700 ºC, $V_{g,crossing}$ (~ -2.0 V) differs from $V_{g,min}$ (~ -1.0 V) in contrast to the results at 800 ºC and 900 ºC showing that $V_{g,crossing} \sim V_{g,min}$. According to the model of Adam *et al.*[31] at low carrier density close to the Dirac point, the potential fluctuations induced by randomly distributed charged impurities break the graphene system up into spatially inhomogeneous puddles of electrons and holes. In this scenario, the conductance minimum is demonstrated to occur at the gate voltage where the average potential is zero instead of the one at the charge-neutral Dirac point. The formation of electron-hole puddles has been evidenced by using scanning probe technique to directly measure the potential fluctuations in graphene.[41] Transport measurements in support of this argument have already been visible in potassium doped graphene.[30] Therefore we believe that the existence of inhomogeneous electron-hole puddles induced by ionized nitrogen impurities contributes to the observed discrepancy between $V_{g,crossing}$ and $V_{g,min}$. The following analysis of $T$-dependent data would further support this argument. Of particular interest is the observation of enhanced electron-hole transport asymmetry in N-doped graphene sheet at the growth temperature of 700 ºC as shown in Fig. 7c compared to those of 800 ºC and 900 ºC presented in Figs. 7b and 7c. The asymmetry feature can be quantified by the electron-hole mobility ratio $\mu_e/\mu_h$ which is obtained by fitting the measured $G(V_g)$ to Eq. (1) and is estimated to be 0.04, 1.04, and 0.73 for the samples grown at 700 ºC, 800 ºC, and 900 ºC, respectively. In other words, the hole mobility can become 25 times larger than the electron one, indicating that electrons scatter off charged impurities much more effectively than holes do, when the grown temperature is reduced to 700 ºC with highest doping concentration in our experiments. In addition, as shown in Fig. 7c for N-doped graphene grown at 700ºC, the n-type behavior with the conductance minimum at $V_g$ = -0.6 V can sustain even in air environment, which has potential applications in graphene-based nanoelectronics. The reduced magnitude of $V_{g,min}$ is mostly due to p-doping by physisorbed molecular oxygen. Based on Novikov's calculation demonstrating that the attractive scattering is stronger than the replusive one for Dirac fermions in graphene,[29] the dominant scatteirng centers in our N-doped graphene sheets are thereby expected to be positive ions. Given the oberved crossover from p-type to n-type behavior and the enhanced transport asymmetry with increasing N doping concentration by decreasing the growth temperature, charged-impurity scattering from positively ionized N atoms as donors may govern the transport behavior in N-doped graphene sheets. However the observed n-type behavior in our N-doped graphene sheets dominated by pyridinic and pyrrolic N doping (see Table 1) seems to be in conflict with the DFT calculations[13, 24-26] which show that pyridinic and pyrrolic N provide p-doping effects. The seminal work of Schiros *et al*. which connects dopant bond type with electronic structure in N-doped graphene[24] may give some insight into this disagreement. The N atom has five valence electrons. The pyridinic N uses three of them to form two σ and one π bonds with carbon neighbors and the remaining two electrons to form a localized electron lone pair. If the formation of this electron lone pair is unfavorable, the additional charge from N would be forced to go to the conduction band, giving rise to n-type doping. For instance, the graphene sheet predominantly doped with graphitic N exhibits strong n-doping effect, in which the formation of electron lone pair is almost prohibited. More importantly, Schiros *et al*.[24] show that pyridinic N can behave as n-type dopant when it is hydrogenated [pyridinic (+H)] even though the charge transfer per N atom is much smaller than the graphitic case. Similar arguments can be valid for pyrrolic N as well. As the growth of N-doped graphene sheets in our experiments proceeds through recombination of triazine molecules, having pyridinic or pyrrolic N bonded with other atoms such as hydrogen mentioned above is naturally expected, which may be a key factor for understanding the observed n-type behavior. It is worth noting that Jin *et al.*[15] recently reported the synthesis of monolayer N-doped graphene by CVD using pyridine as the sole source of both carbon and nitrogen, and showed that the N atoms doped in the graphene are mainly in the form of pyridinic N. Correspondingly, Jin *et al.* also observed n–type behavior with the neutrality point at -10 V. In our experiments, the disorder-related $V_{g,crossing}$ moves from -1.1 V to -2 V (see Table 2) when the atomic percentage of graphitic N shows a decrease (see Table 1) with reducing the growth temperature from 800 ºC to 700 ºC. It is interesting to see from Table 1 that both pyridinic and pyrrolic N contents (atomic percentages) increase with lowering the temperature of CVD growth. Therefore we believe that the peculiar n-type behavior in N-doped graphene sheets is as well relevant to pyridinic and pyrrolic N instead of being totally attributed to graphitic N. Here we should also mention that the p-type characteristics observed in the sample grown at 900 ºC may be due to physisorbed molecular oxygen, residual PMMA used during transfer process, or doping mechanism. Therefore further investigations on the behavior of nitrogen doping at different growth temperatures are required, which paves the way for controllable doping of graphene.

Further support for the existence of electron-hole puddles stems from the analysis of temperature-dependent conductance $G(T)$ at the gate voltage near $V_{g,min}$, which is shown in the insets of Figs. 7a-7c at the corresponding growth temperatures. All the samples exhibit insulating behavior in the sense that $G$ increases with increasing $T$. Several competing mechanisms such as localization[42-44] and thermal excitation effects[20, 45] can lead to such behavior. Figure 8a shows that $G$ increases quadratically with increasing $T$ for all the samples grown at different temperatures, reminiscent of the recent proposed model concerning disorder-induced $T$-dependent transport in graphene.[46] In the presence of inhomogeneous electron-hole puddles induced by charged impurities, it is predicted that direct thermal excitation of carriers from the valence band to the conduction band or thermal activation across the potential fluctuations associated with these puddles would play a key role in the insulating behavior of $G(T)$. At high $T$ ($k_BT>s$, where $k_B$ is Boltzmann constant and $s$ represents the root-mean-square potential fluctuations), thermal excitation of carriers dominates the transport and gives a quadratic temperature-dependent term to the conductance near charge neutrality point. At low $T$ ($k_B T < s$), thermal activation across puddles is dominant instead, which gives rise to a linear $T$ dependence of the conductance. In other words, thermal excitation of carriers and thermal activation across the disorder-induced puddles dominate the transport in weak and strong potential fluctuation regimes, respectively.[46] The observed $T^2$ dependence of $G$ as shown in Fig. 8a is ascribed to the direct thermal excitation behavior. To further understand the transport mechanism in graphene with the presence of nitrogen impurities in the carbon network, the electrical measurements were further performed on N-doped graphene sheets transferred onto GaAs substrate, in which the potential fluctuations are believed to be stronger than the one transferred onto $SiO_2$/Si substrate since the undpoed GaAs substrate cannot screen the impurity charges in graphene as well as the $SiO_2$/Si substrate does. Figure 8b shows the results of conductance measurements on the samples grown at three corresponding temperatures with the same CVD process but on the GaAs substrates. It can be seen clearly that $G$ increases linearly with increasing $T$ for all the samples, a signature of activated transport across puddles. Such results confirm the formation of electron-hole puddles induced by ionized nitrogen impurities, which gives a plausible reason for the observed discrepancy between $V_{g,crossing}$ and $V_{g,min}$ in the sample grown at 700 ºC as shown in Fig. 7c. In conventional 2D systems, strong enough disorder results in strong localization of carriers (Anderson localization) and the conductance can become extremely small as the zero-temperature limit is approached.[43, 47] Therefore strong insulating $T$ dependence is expected to occur in a highly disordered system. However for graphene in the presence of only long-range scatterers, carriers can still conduct instead of being localized (owing to the absence of backscattering) even in the zero-temperature limit.[5, 6] On the other hand, with a sufficient amount of short-range disorder, transport properties of graphene would resemble those of conventional 2D systems due to strong intervalley mixing.[48] As learned from XPS studies showing that pyridinic N and pyrrolic N (which cause intervalley scattering) doping are dominant, localization effects may also be important in our N-doped graphene sheets and become increasingly pronounced with enhancing the doping level. Figures 6c and 6d show that $G(T)$ of the sample grown at 700 ºC either on $SiO_2$/Si or GaAs substrate can also be well described by $G(T) = G_0 \exp(-(T_0/T)^{1/3})$ with a prefactor $G_0$ and characteristic temperature $T_0$, which is a signature of variable-range hopping (VRH) transport.[42] For the samples grown at higher temperatures, VRH behavior cannot be found. It is known that VRH occurs as the carriers are strongly localized.[43] Thses findings are consistent with XPS and Raman spctroscopy data showing that the amount of disorder leading to intervalley scattering increases with lowering the growth temperature. Some deviations from $T^2$ and $T$ dependence observed at low $T$ regime as shown in

Figs. 8a and 8b may be due to the fact that some factors are not considered in the puddle model[46] or the influence of localization effects which is still unclear and requires further investigations. Recently, there has been a growing interest in studying the coexistence of electron-hole puddles and localization in disordered graphene.[49-51] This issue is especially important in N-doped graphene sheets dominated by pyridinic and pyrrolic N doping as these two bonding configurations are always accompanied by the appearance of structural defects which tend to localize the transport carriers. In our present study, we show that charged-impurity scattering, electron-hole puddles, and localization effects are essential for understanding the 2D physics in N-doped graphene sheets. Ideally we may compare samples with exactly the same thickness at different growth temperatures. We note that increasing the film thickness should improve the conducting properties. In contrast, increasing doping concentration makes graphene sheets become much more disordered, which gives rise to a stronger temperature dependence of conductance $G(T)$. As shown in Fig. 8a, $G(T)$ for N-doped graphene sheets with different doping level becomes monotonically stronger with increasing doping concentration, which excludes the possibility of significant influence due to layer number variation.

Finally, it is worth mentioning that a lot of theoretical work has focused on the electronic structure of few-layer graphene and demonstrate that the nature of charge carriers is strongly related to the number of layers and the stacking geometry.[52-54] Under specific conditions, few-layer graphene can behave as single-layer graphene, mixed carriers semimetal with overlapping conduction and valence bands, or semiconductor with an energy gap.[52] Therefore our experimental finding on N-doped graphene sheets, which can be at least qualitatively described by the theory derived for single-layer graphene in the presence of charged impurities in its environment rather than in itself, may stimulate further experimental and theoretical investigations of doping effect on few-layer graphene system. Our experimental results on XPS, Raman spectroscopy, and transport measurements show that the electrical properties of graphene would be significantly altered as the nitrogen impurities are incorporated into its carbon lattice.

**Conclusion**

In conclusion, we have successfully synthesized nitrogen-doped graphene sheets by CVD using organic molecule 1, 3, 5-triazine as a solid precursor. With lowering the CVD growth temperature, the concentration of incorporated nitrogen impurities, where pyridinic and pyrrolic N are dominant, increases. In addition, the nitrogen-doped graphene sheet exhibits a crossover from p-type to n-type behavior and a strong enhancement of electron-hole transport asymmetry with increasing doping concentration, which indicates that scattering by ionized nitrogen impurities plays a significant role in determining the carrier transport behavior. The n-type behavior can still be seen in ambient air environment, which is of great technological importance for manipulating graphene in nanoelectronics. Furthermore, by studying the insulating temperature dependence of conductance, we demonstrate that electron-hole puddles induced by the potential of charged impurities should be considered to understand the transport features near the conductance minimum.

## Methods

**Synthesis of N-doped graphene sheets.** In this study, N-doped graphene was prepared by low pressure CVD of 1,3,5-triazine (abbreviated as triazine) $(HCN)_3$ molecules on Cu foils. In our CVD system, there is an additional quartz tube, which is connected to the main one, for hosting the triazine. A 25-µm-thick Cu foil was placed inside the horizontal quartz tube with $H_2$ and Ar flows both at 300 sccm (standard cubic centimeters per minutes) and a total pressure of 500 torr. The furnace was heated up to 1000 ºC at a ramping rate of 30 ºC/min and then maintained at 1000 ºC for 30 minutes. After annealing, the furnace was cooled down to 500 ºC and evacuated for a few minutes. At the same time, the triazine was thermally evaporated at 150 ºC for 15 minutes before feeding it into the main tube maintained at 500 ºC. The gaseous triazine was then released into the main tube and absorbed onto the Cu surface for 20 minutes. Then the valve, separating the main tube from the one for triazine, was closed and the sample was rapidly retreated back from the center of the main tube. After that, the main tube was heated up to the desired growth temperature (900 ºC, 800 ºC, and 700 ºC) under a 10 sccm $H_2$ flow and a Ar 100 sccm flow with a total pressure maintained at 5 torr. Once the temperature was reached, the sample was moved back to the center for 10 minutes, an important procedure to avoid the aggregation of the Cu catalyst before the growth of doped graphene. At high enough temperatures, the absorbed triazine molecules were dissociated and then recombined into the N-doped graphene by the aid of Cu.

**Device fabrication and transport measurements.** For electrical measurements, the N-doped graphene sheets were transferred onto both 300 nm $SiO_2$/Si and GaAs substrates. The first step of the transfer is to spin-coat a thin layer of polymethyl-methacrylate (PMMA) onto the as-grown N-doped graphene sheets as a support to reinforce the graphene structure. The PMMA/graphene/Cu film was baked for 15 minutes at 130 ºC to solidify the PMMA. Then, Cu foil was etched by an aqueous solution of iron nitrate ($FeCl_3$), yielding a PMMA/graphene film. After removal of the Cu foil, the PMMA/graphene film was carefully rinsed in the deionized (DI) water. Then it was lifted from the solution and transferred onto the desired substrate. Finally, the PMMA layer was dissolved in acetone, which was removed later by isopropyl alcohol (IPA). In the following, Au was thermally deposited on the transferred N-doped graphene sheets in a two terminal configuration to form source/drain contacts. Transport measurements were performed in a closed-cycle system using two-probe techniques over the temperature range of $30\ K \leq T \leq 300\ K$ for the samples on $SiO_2$/Si substrates and of $40\ K \leq T \leq 290\ K$ for those on GaAs substrates. The applied current for the transport measurements was 100 nA, 5 µA, and 2 µA for the samples grown at 900 ºC, 800 ºC, 700 ºC, respectively. For the sake of comparison with theoretical calculations, we present the results by the form of conductance $G = 1/R$, the inverse of the obtained resistance. Moreover, the range of applied gate voltage is from 0 to 9 V, from -2.6 V to 2.8 V, and from -2.7 V to 1.6 V for the samples grown at 900 ºC, 800 ºC, 700 ºC, respectively. The small voltage range around the Dirac point was chosen to avoid the back-gate leakage current in our large-scale devices.

**Characterizations.** The AFM images were obtained in a Veeco Dimension-Icon system. The surface morphology of the samples was examined with a scanning electron microscope (SEM, FEI VS 600). Transmission electron microscope (Tecnai G2 F20 FEI operated at 200 kV) at the Department of Physics, National Taiwan University was used to obtain information of the microstructures. Raman spectra were measured in a NT-MDT confocal Raman microscopic system with a 473 nm excitation laser. XPS measurements were carried out on a Kratos AXIS spectrometer (UK) with a monochromatic Al Ka X-ray radiation at 1486.71 eV. The AFM images and Raman spectra were taken from N-doped graphene sheets transferred onto $SiO_2$/Si substrates while XPS spectra were taken from as-grown ones on Cu foil.

**Acknowledgments**
This work was funded by the NSC, Taiwan and National Taiwan University (grant no: 102R7552-2, 102R890932, 101R7800-3 and 102R0044). We would like to thank Xin-Quan Zhang for performing the SEM measurements.

**Supporting Information Available**: Detailed results of Lorentzian fits to the Raman spectra for N-doped graphene sheets at different synthesis temperatures. This material is available free of charge *via* the Internet at http://pubs.acs.org.

Figure Captions

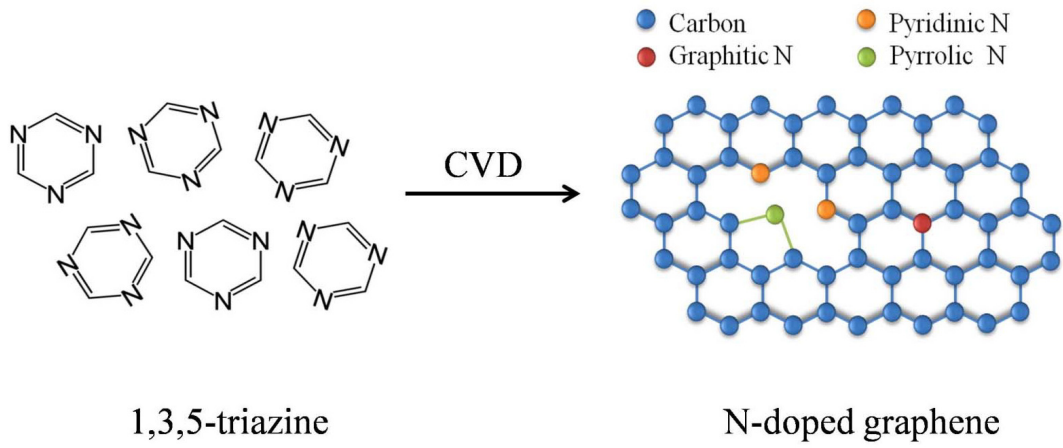

1,3,5-triazine    N-doped graphene

Figure 1 Schematic structures of 1, 3, 5-triazine molecule and N-doped graphene.

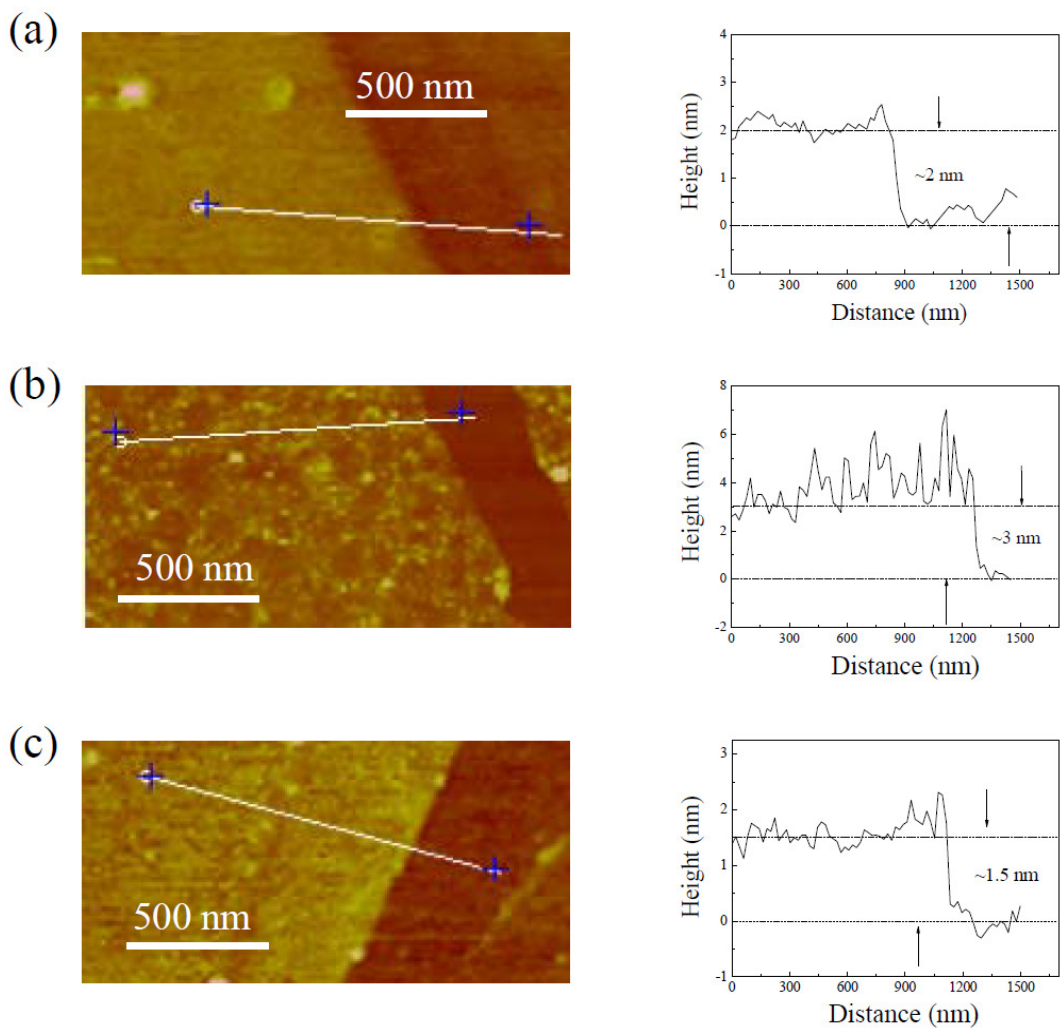

Figure 2 AFM images and cross-section profiles of N-doped graphene sheets grown at (a) 900ºC, (b) 800 ºC, and (c) 700 ºC.

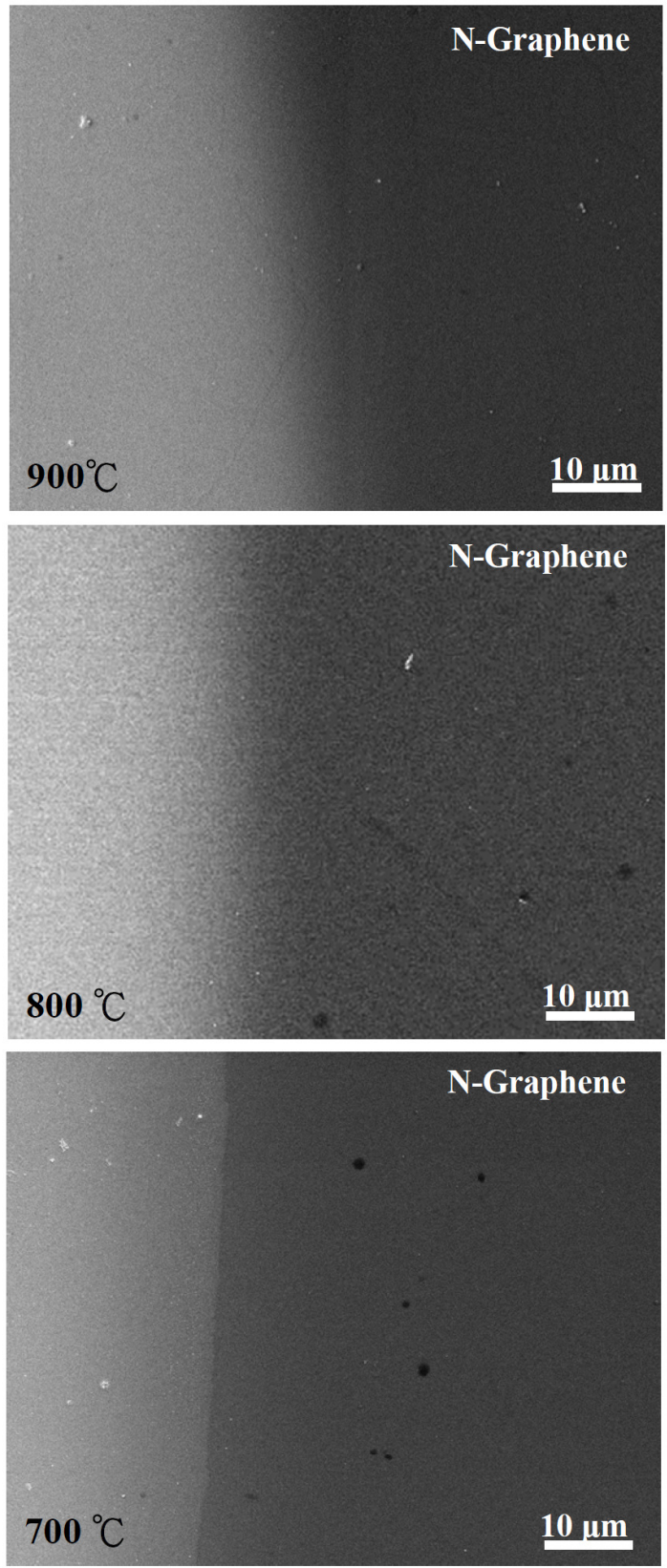

Figure 3 SEM images of N-doped graphene sheets grown at (a) 900ºC, (b) 800 ºC, and (c) 700 ºC.

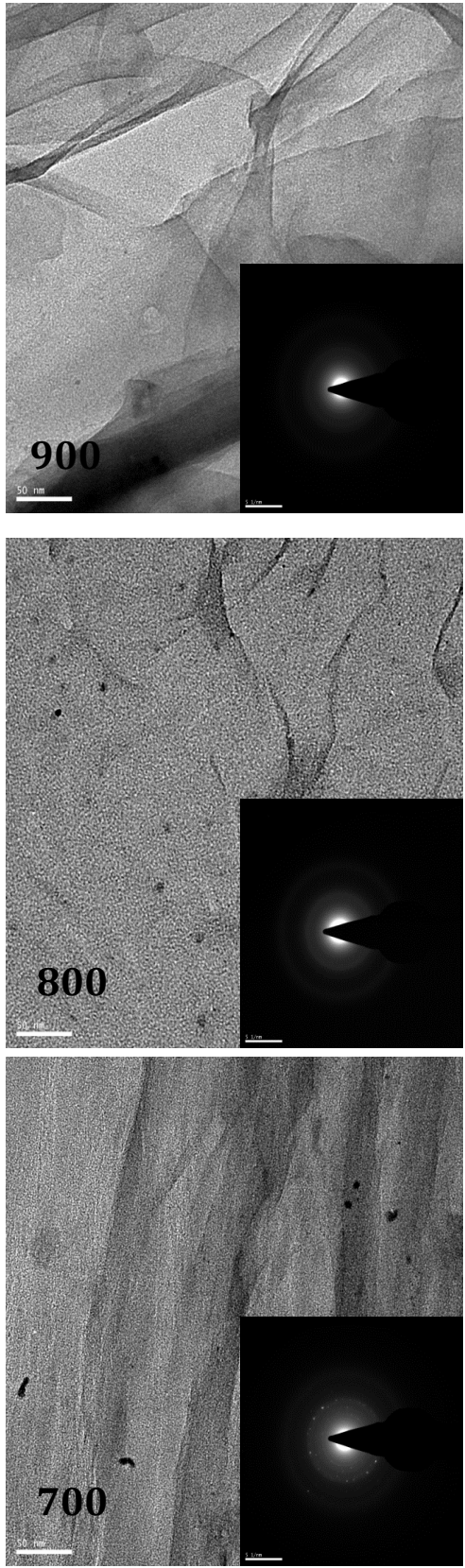

Figure 4 TEM images with the insets showing the SAED patterns of N-doped graphene sheets grown at (a) 900 ºC, (b) 800 ºC, and (c) 700 ºC.

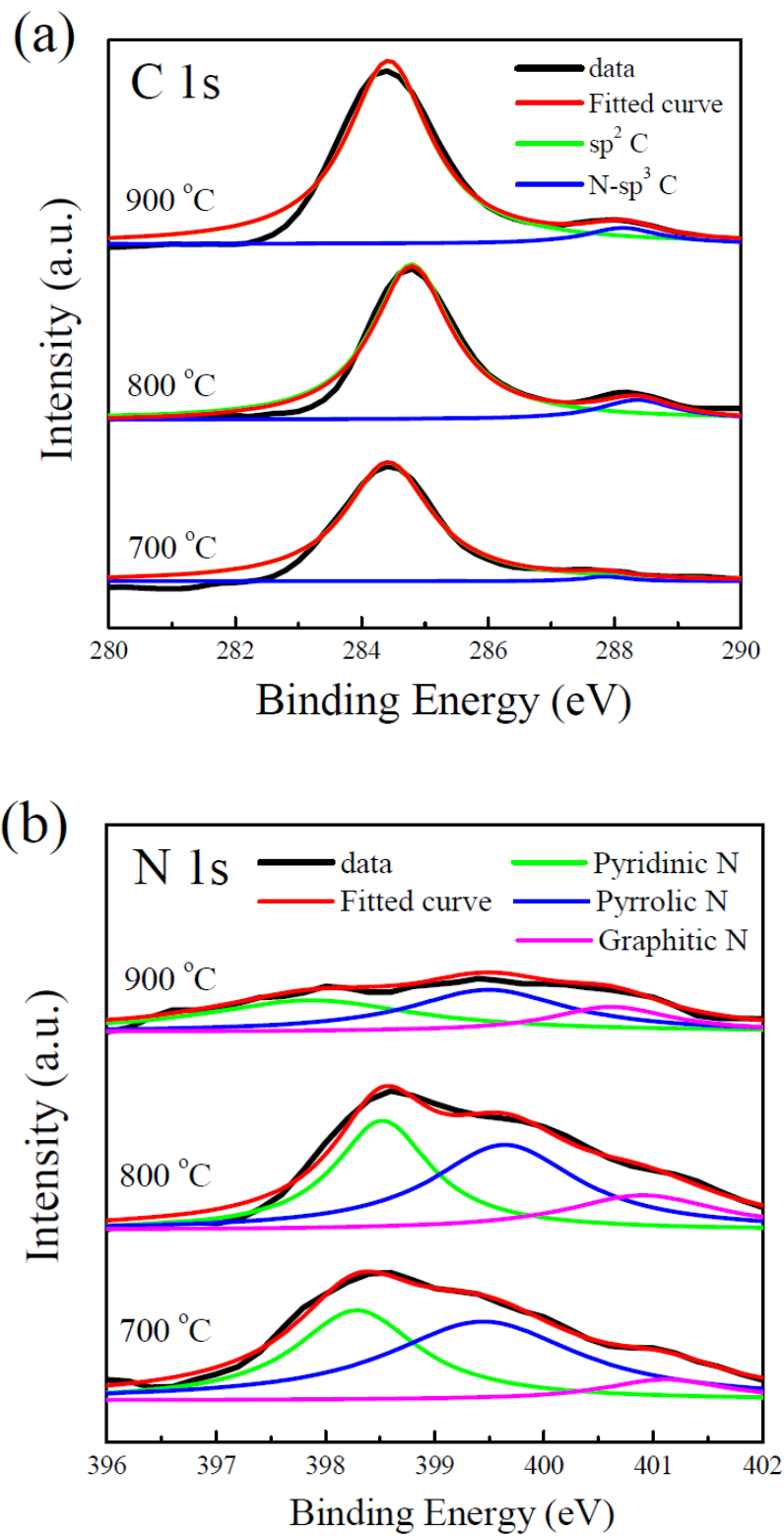

Figure 5 (a) XPS C 1s and (b) N 1s spectra of N-doped graphene sheets at three corresponding growth temperatures.

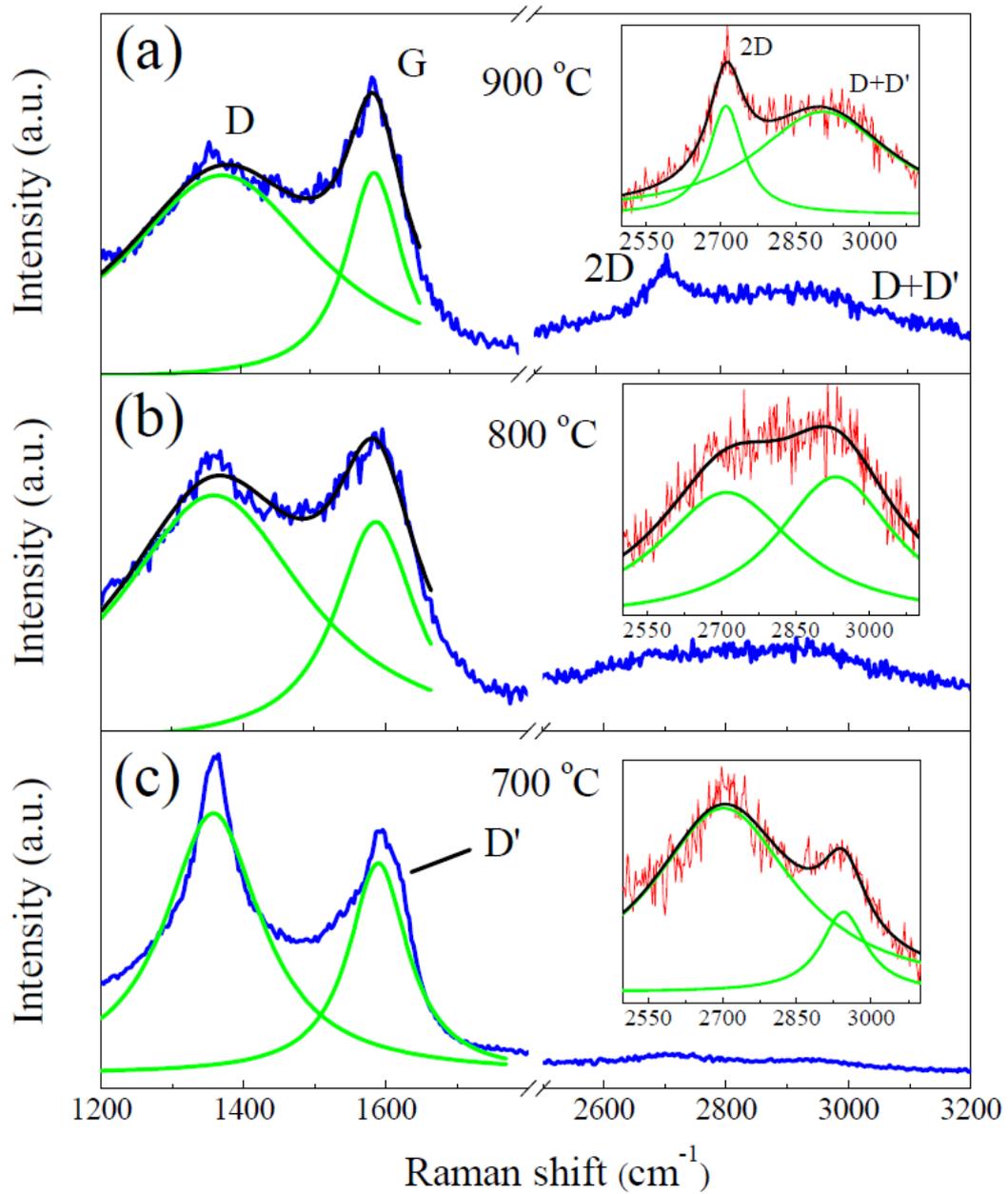

Figure 6 Raman spectra and their Lorentzian fits of N-doped graphene sheets at three corresponding growth temperatures.

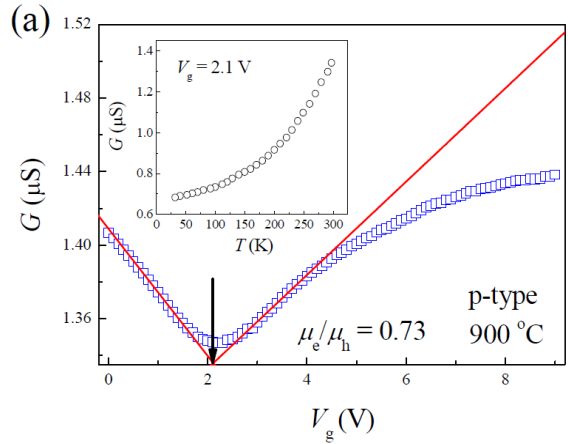

Figure 7 (a)

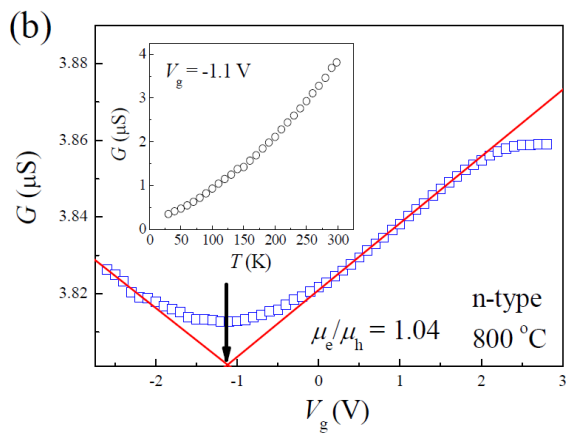

Figure 7 (b)

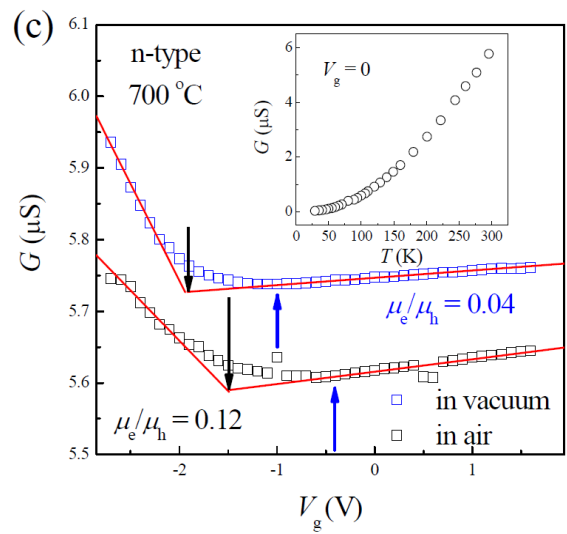

Figure 7 (c)

Figure 7 Conductance $G$ as a function of gate voltage $V_g$ for N-doped graphene sheets grown at 900 °C (a), 800 °C (b), and 700 °C (c). The red lines correspond to the fits to Eq. (1). The insets show $G$ as a function of temperature $T$ at the gate voltage near the point where the conductance minimum occurs.

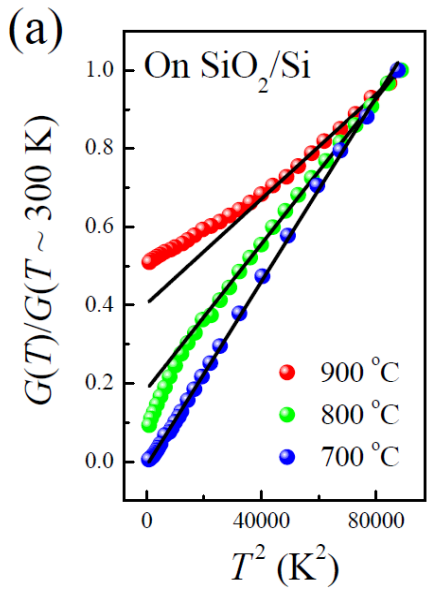
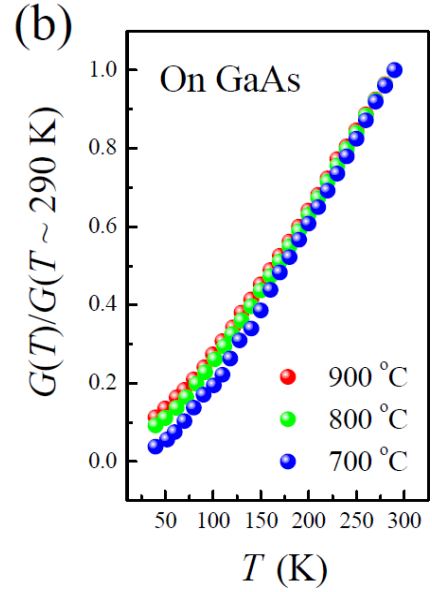
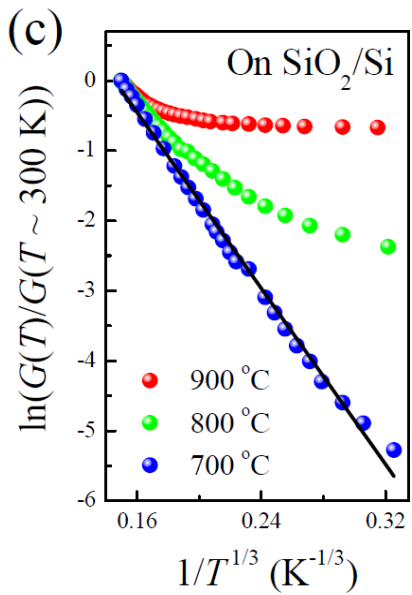
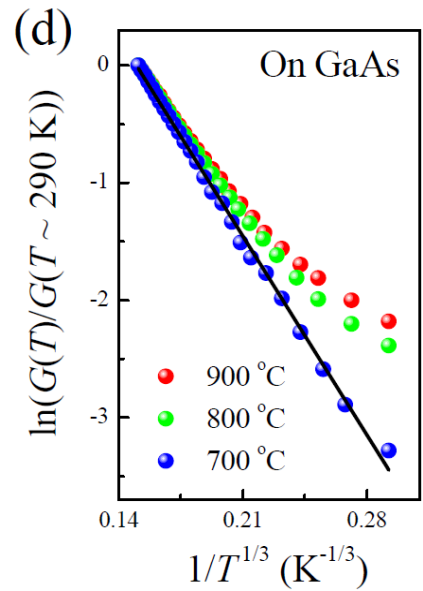

Figure 8 (a) $G$ as a function of $T^2$ for the N-doped graphene sheets transferred onto SiO$_2$/Si substrates. (b) $G$ as a function of $T$ for those transferred onto GaAs substrates. (c) and (d) show ln$G$ versus $T^{-1/3}$ for the samples transferred onto SiO$_2$/Si and GaAs substrates, respectively. For comparison between different growth temperatures, the conductance is normalized to its value at the highest measurement temperature.

**TABLE 1. Fitted results of XPS spectra**

| Growth temp. | sp$^2$ C BE (eV) | N-sp$^3$ C BE (eV) | Pyridinic N BE (eV) | Pyridinic N ratio (%) | Pyrrolic N BE (eV) | Pyrrolic N ratio (%) | Graphitic N BE (eV) | Graphitic N ratio (%) | N/C$^a$ (at. %) |
|---|---|---|---|---|---|---|---|---|---|
| 900 ºC | 284.4 | 288.1 | 397.9 | 41 | 399.5 | 41 | 400.6 | 18 | 2.1 |
| 800 ºC | 284.8 | 288.3 | 398.5 | 37 | 399.6 | 44 | 400.9 | 19 | 4.4 |
| 700 ºC | 284.4 | 287.9 | 398.3 | 39 | 399.4 | 52 | 401.1 | 9 | 5.6 |

$^a$The N/C atomic ratio was calculated by taking the ratio of the total area under the N 1s spectrum to that under the C 1s one.

**TABLE 2. Characteristic values obtained from Raman spectra and electrical measurements**

| | 900 ºC | 800 ºC | 700 ºC |
|---|---|---|---|
| $I_D/I_G$ | 0.99 | 1.12 | 1.24 |
| $I_{2D}/I_G$ | 0.23 | 0.14 | 0.07 |
| $\mu_e$ (cm$^2$/Vs) | 1.10 | 0.76 | 0.45 |
| $\mu_h$ (cm$^2$/Vs) | 1.50 | 0.73 | 11.7 |
| $\mu_e/\mu_h$ | 0.73 | 1.04 | 0.04 |